\documentstyle[pra,aps,multicol,epsfig]{revtex}
\begin{document}

\title{Multidimensional solitons in a low-dimensional periodic potential}
\author{Bakhtiyor B. Baizakov$^1$, Boris A. Malomed$^2$, and Mario Salerno$^1 $}

\address{$^1$ Dipartimento di Fisica "E.R. Caianiello" \\
and Istituto Nazionale di Fisica della Materia (INFM), \\
Universit\'a di Salerno, I-84081 Baronissi (SA), Italy, \\
$^2$ Department of Interdisciplinary Studies, Faculty of Engineering, \\
Tel Aviv University, Tel Aviv 69978, Israel}
\date{\today }

\maketitle

\begin{abstract}
Using the variational approximation(VA) and direct simulations, we find
stable 2D and 3D solitons in the self-attractive Gross-Pitaevskii equation
(GPE) with a potential which is uniform in one direction ($z$) and periodic
in the others (but the quasi-1D potentials cannot stabilize 3D solitons).
The family of solitons includes single- and multi-peaked ones. The results
apply to Bose-Einstein condensates (BECs) in optical lattices (OLs), and to
spatial or spatiotemporal solitons in layered optical media. This is the
first prediction of {\em mobile} 2D and 3D solitons in BECs, as they keep
mobility along $z$. Head-on collisions of in-phase solitons lead to their
fusion into a collapsing pulse. Solitons colliding in adjacent OL-induced
channels may form a bound state (BS), which then relaxes to a stable
asymmetric form. An initially unstable soliton splits into a three-soliton
BS. Localized states in the self-repulsive GPE with the low-dimensional OL
are found too.
\end{abstract}

PACS numbers: 03.75.Fi, 05.30.Jp, 05.45.-a

\begin{multicols}{2}

{\it Introduction.} Solitons in multidimensional nonlinear Schr\"{o}dinger
equations (NLSEs) with a periodic potential have recently attracted
considerable interest. In particular, self-trapping of spatial beams in
nonlinear photonic crystals is described by a 2D equation. In this case,
despite the possibility of collapse \cite{Berge'}, simulations reveal robust
2D solitons in the self-focusing model \cite{PhotCryst}. A similar medium
can be created by a grid of laser beams illuminating a photorefractive
sample \cite{Moti}.

Similar 2D and 3D models with a periodic potential describe a Bose-Einstein
condensate (BEC) trapped in an optical lattice (OL) \cite{OL}; this
application is relevant, as experimental techniques for loading BECs into
multidimensional OLs were recently developed \cite{loadingMott}. Stable
solitons can be supported by an OL even in self-repulsive BECs
\cite{bks2002,ok2003}. In the case of self-attraction, 2D and 3D solitons
(including 2D vortices) are stable in the self-focusing model with the OL
potential \cite{bms2003}, despite the possibility of the collapse.

An issue of direct physical relevance, and a subject of the present work,
are multidimensional solitons in media equipped with periodic potentials of
a {\em lower dimension}, viz., quasi-1D (Q1D) and Q2D lattices in the 2D and
3D cases, respectively. In optics, the 2D equation in the spatial domain
governs the beam propagation in a layered bulk medium along the layers,
which is an extension of a 1D multichannel system introduced in Ref.
\cite{Wang}, where the potential was induced by transverse modulation of the
refractive index (RI). In the temporal domain, the 2D and 3D equations
govern, respectively, the longitudinal propagation of spatiotemporal optical
solitons\ in a layered planar waveguide, or in a bulk medium with the RI
periodically modulated in both transverse directions. The 2D and 3D cases
directly apply to BECs loaded in a Q1D or Q2D lattice. The physical
significance of this setting is two-fold: first, in the experiment it is
much easier to create a lower-dimensional lattice than a full-dimensional
one, both in BECs and in optics, hence the use of such lattices is the most
straightforward way to create multidimensional solitons in these media.
Second, the solitons created this way can freely move in the unconfined
direction, which also suggests a possibility to study their collisions, and
to look for their bound states (BSs). As yet, no other way to create
multidimensional {\em mobile} solitons in BECs and their BSs has been
proposed. In optics, solitons of this type suggest new applications. Indeed,
in an optical medium with the full-dimensional periodic potential, transfer
of a trapped beam from one position to another is difficult, as the
necessary external push strongly disturbs the beam \cite{Wang}. In the
lower-dimensional potential, the beam can slide along the free direction,
making the transfer easy. In BECs confined by a low-dimensional OL,
matter-wave solitons can be driven in the free direction by a weak laser
beam.

{\it The model and VA.} The normalized form of the self-focusing 2D NLSE
with a Q1D periodic potential of the strength $\varepsilon $ is
\begin{equation}
iu_{t}+u_{xx}+u_{yy}+[\varepsilon \cos (2x)+\chi |u|^{2}]u=0,  \label{gpe}
\end{equation}
where $\chi =\pm 1$ in the case of the self-attraction/repulsion. In BECs, 
$t $ is time, while in optics it is the propagation distance. For BECs or
spatial optical solitons, $x$ and $y$ are transverse coordinates; for
spatiotemporal optical solitons in a 2D waveguide with anomalous chromatic
dispersion, $y$ is a temporal variable. The 3D version of Eq. (\ref{gpe}) is 
\begin{equation}
iu_{t}+u_{xx}+u_{yy}+u_{zz}+\left\{ \varepsilon \left[ \cos (2x)+\cos (2y)
\right] +\chi |u|^{2}\right\} u=0.  \label{3D}
\end{equation}
Equations (\ref{gpe}) and (\ref{3D}) conserve the Hamiltonian, the norm 
$N=\int \left\vert u({\bf r})\right\vert ^{2}d{\bf r}\,$ (the number of atoms
in the BEC, or total power/energy of the spatial/spatiotemporal optical
soliton), and the momentum along the free direction. The equations are
normalized so that the period of the potential is $\pi $, the free
parameters being $\varepsilon $ and $N$.

\begin{figure}[htbp]
\centerline{\includegraphics[width=8.0cm,height=4.0cm,clip]{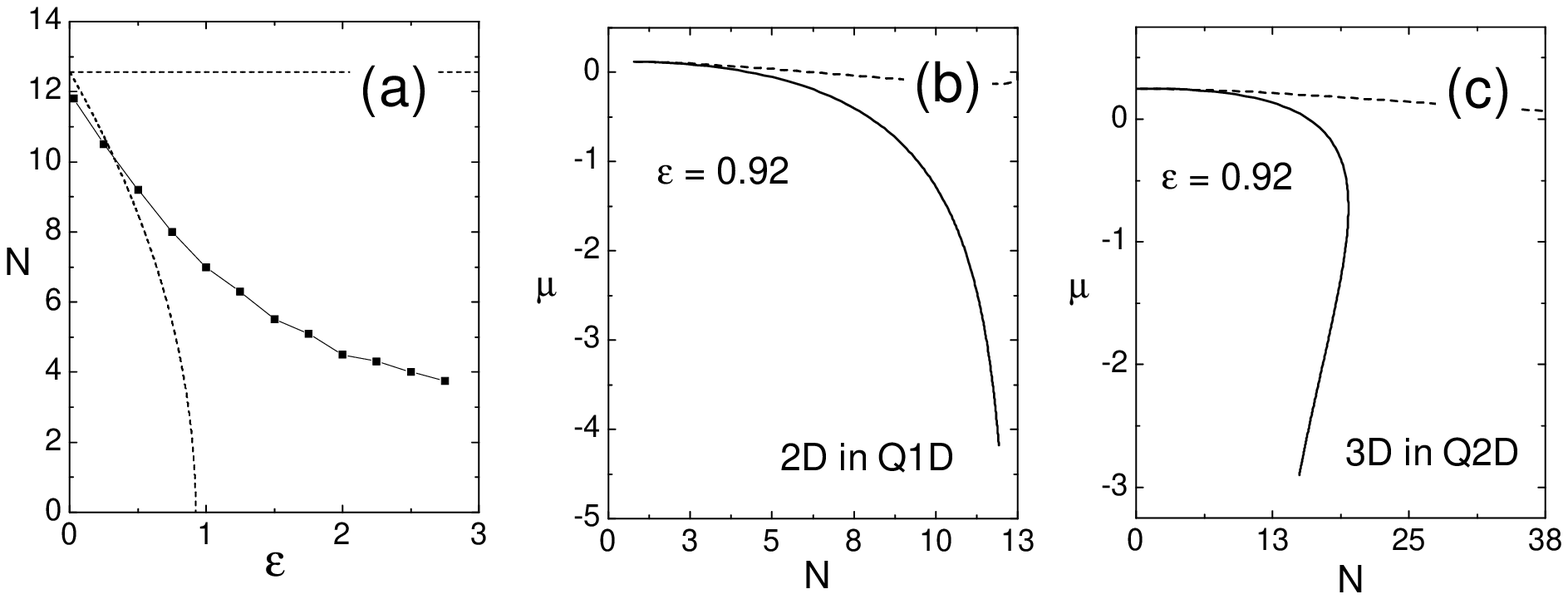}}
\caption{The numerically found (joined squares) and VA-predicted (dashed
lines) existence limits for stable 2D solitons (a). VA-predicted dependences
$\protect\mu(N)$ for the 2D (b) and 3D (c) solitons, which determine their
VK stability.}
\label{f1}
\end{figure}

Stationary solutions to Eq. (\ref{gpe}) are looked for as
$u(x,y,t)=U(x,y)\exp (-i\mu t)$, with a chemical potential $\mu $
(propagation constant, in optics), which yields 
\begin{equation}
\mu U+U_{xx}+U_{yy}+\left[ \varepsilon \cos (2x)+\chi U^{2}\right] U=0,
\label{stat2D}
\end{equation}
and similar in the 3D case. To apply the VA (variational approximation) to
Eq. (\ref{stat2D}), we adopt the {\it ansatz} $U=A\exp \left[ -\left(
ax^{2}+by^{2}\right) /2\right] $, or $U=A\exp \left[ -\left(
a(x^{2}+y^{2})+bz^{2}\right) /2\right] $ in the 3D case, with the norms $N_{2
{\rm D}}=\pi A^{2}/\sqrt{ab}$, and $N_{3{\rm D}}=\pi ^{3/2}A^{2}/(a\sqrt{b})$.
Following the procedure of the application of VA to multidimensional
solitons \cite{malomed}, we derive variational equations from the Lagrangian
of Eq. (\ref{stat2D}). In the 2D case, they are ($\aleph \equiv e^{-1/a}$) 
\begin{equation}
N\chi =\left( 4\pi /a\right) \sqrt{a^{2}-2\varepsilon \aleph },~\mu
=-a-\varepsilon \aleph (1-3/a),  \label{va3}
\end{equation}
and in the 3D case [in both cases, $b=\left( a^{2}-2\varepsilon \aleph
\right) /a$], 
\begin{equation}
N\chi =2\left( 2\pi /a\right) ^{3/2}\sqrt{a^{2}-2\varepsilon \aleph },~\mu
=-a/2-\varepsilon \aleph \left( 2-3/a\right) .  \label{VA3D}
\end{equation}
Evidently, solutions are possible only if $\chi >0$ (self-attraction). The
2D solutions exist in the interval
\begin{equation}
N_{\min }\equiv 4\pi \sqrt{1-\varepsilon /\varepsilon _{{\rm cr}}}<N<N_{\max
}\equiv 4\pi  \label{minmax}
\end{equation}
[see Fig. \ref{f1}(a)], if $\varepsilon <\varepsilon _{{\rm cr}}\equiv
e^{2}/8\approx \allowbreak 0.924$ ($N_{\min }=0$ if $\varepsilon
>\varepsilon _{{\rm cr}}$), while 3D solutions can be found for any $N$. The
fixed lattice period rules out establishing general scaling relations for
these soliton families; in the case of very large $\varepsilon $, the
potential valleys become isolated, then one is actually dealing with
cigar-shaped \cite{1Dsol} Q1D solitons, that obey the usual 1D NLSE scaling.

The solution families which definitely meet the condition $d\mu /dN<0$ are
expected to be stable pursuant to the {\it Vakhitov-Kolokolov} (VK)
criterion \cite{vk}, while ones with $d\mu /dN$ positive or nearly zero
should be unstable. It follows from Eqs. (\ref{va3}) and (\ref{VA3D}) and is
evident in Figs. \ref{f1}(b,c) that the 2D solitons are stable in the
existence interval (\ref{minmax}). At small $\varepsilon $, the 3D solutions
are unstable for any $N$. Equation (\ref{VA3D}) predicts that a stability
interval of a width $\Delta N\sim \sqrt{\varepsilon -\varepsilon _{0}}$
appears around $N_{0}\approx 10.1\pi ^{3/2}$ if $\varepsilon $ exceeds
$\varepsilon _{0}\approx 0.242$.


\begin{figure}[htbp]
\centerline{\includegraphics[width=8.0cm,height=8.0cm,clip]{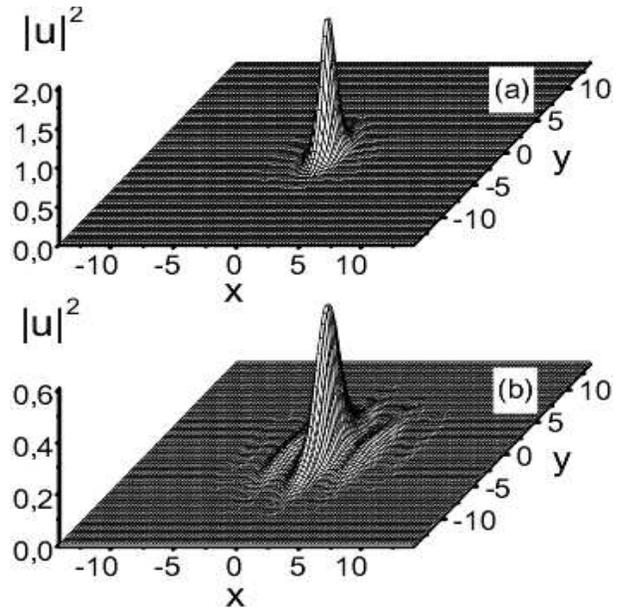}}
\vspace*{0.5cm}
\caption{Examples of single- and multi-peaked stable 2D solitons in the Q1D
potential ($\protect\varepsilon = 2$). The norm of the soliton is
$N=2\protect\pi$ (a) and $N=1.5 \protect\pi$ (b). }
\label{f2}
\end{figure}

\begin{figure}[htbp]
\centerline{\includegraphics[width=8.0cm,height=8.0cm,clip]{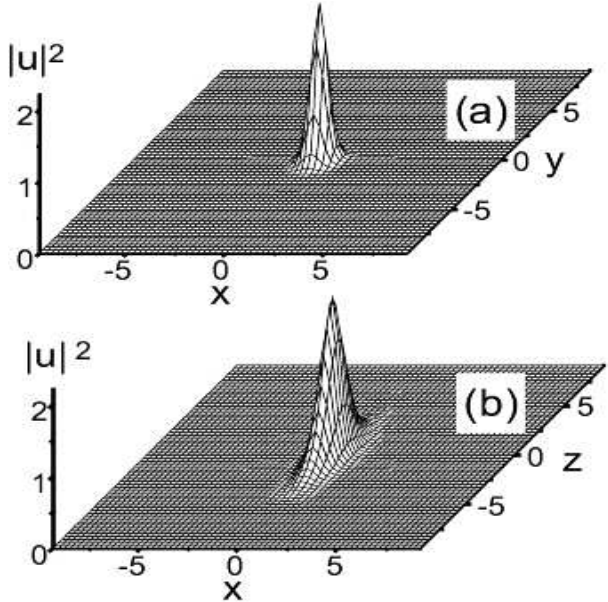}}
\vspace*{0.5cm}
\caption{A stable single-peaked 3D soliton in the Q2D potential
($N=2\pi$, $\protect\varepsilon =5.0$) is shown through its $z=0$ (a)
and $y=0$ (b) cross sections.}
\label{f3}
\end{figure}

{\it Numerical results.} Solitons are generated by the imaginary-time
evolution method \cite{chiofalo}, starting with the VA-predicted waveforms.
Stability is verified by direct simulations of perturbed solitons in real
time. Typical examples of the thus found stable 2D and 3D solitons are
displayed in Fig. \ref{f2} and \ref{f3}, the 3D ones being nearly isotropic
in the $\left( x,y\right) $ plane and elongated in the free direction $z$.
As is seen in Fig. \ref{f1}(a), the existence limits for the stable 2D
solitons are close to the VA prediction, unless $N$ is small; in that case,
the discrepancy is due to the fact that the soliton develops a multi-peaked
structure [Fig. \ref{f2}(b)], which is not accounted for by the simple
ansatz adopted above. The ansatz can be generalized for this case, setting
$U=\left( A+B\,\cos x\right) \exp \left[ -\left( ax^{2}+by^{2}\right) /2
\right] $, but the final result is then quite messy. A numerically found
stability region for the 3D solitons is also quite close to the VA
prediction, and multi-peaked 3D solitons relate to their single-peaked
counterparts (Fig. \ref{f3}) the same way as Fig. \ref{f2}(b) to Fig.
\ref{f2}(a).

The VA predicts that, in the 3D case with the Q1D (rather than Q2D) lattice,
all the solitons are VK-unstable. Accordingly, simulations {\em never}
produced stable solitons in this case. This feature can be explained by the
fact that, in the free 2D subspace, the soliton is essentially tantamount to
the unstable {\it Townes soliton} \cite{Berge'}. In the 2D NLSE with the 2D
lattice, stable solitons with intrinsic vorticity were also found in Ref.
\cite{bms2003}. Vortices were found in the present 2D model too, but they
are always unstable.

We have also investigated the case of self-repulsion, $\chi =-1$, when the
low-dimensional lattice potential cannot support a completely localized
pulse. However, adding a usual parabolic trap readily gives rise to stable
states, which are solitons across the lattice and Thomas-Fermi states along
the free direction. It is noteworthy that, with $\chi <0$, these states
assume a multi-peaked or single-peaked shape under the action of a weaker or
stronger OL potential, respectively, which is {\em opposite} to what was
reported above for $\chi >0$, cf. Fig. \ref{f2}. An explanation is that, in
the case of the self-repulsion and $\varepsilon =0$, no solitons exist (even
unstable ones, like the above-mentioned Townes soliton).

The influence of the parabolic trap was checked too in the case of
$\chi =+1$. If the 2D or 3D soliton is displaced from the central position, it
performs harmonic oscillations along the free direction, completely
preserving its integrity. Actually, the {\em mobility} in the free direction
is the most essential difference of the present multidimensional solitons
from those predicted in other BEC models \cite{bks2002,ok2003,bms2003}. This
suggests new possibilities, such as collisions between moving solitons.

Simulations demonstrate that 2D and 3D in-phase solitons which collide
head-on with velocities $\pm v$, which are below a critical value
$v_{{\rm c}}$, merge into a single pulse, whose norm exceeds the critical
value $N_{\max }$ [see Eq. (\ref{minmax})], hence it quickly collapses.
If $v>v_{{\rm c}}$, the solitons pass through each other (for instance,
$v_{{\rm c}}=8.5$ for $\varepsilon =2.0$, when the norm of
each 2D soliton is $N=2.5\pi$), which is explained by the fact that
the collision time, $\sim 1/v$, is
then smaller than the collapse time, $\sim 1/N$. Colliding
$\pi $-out-of-phase solitons always bounce back, and two such solitons, placed
inside the trap, perform stable oscillations with periodic elastic
collisions.

\begin{figure}[htbp]
\centerline{\includegraphics[width=8.0cm,height=8.0cm,clip]{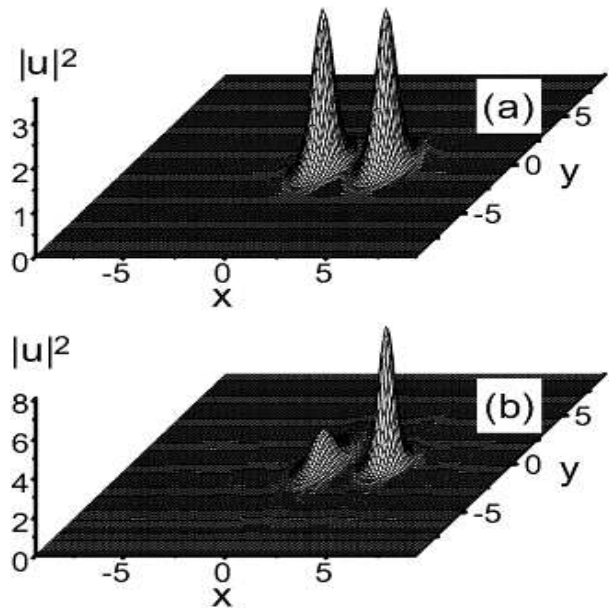}}
\vspace*{0.5cm}
\caption{A symmetric bound state (BS) of 2D in-phase solitons formed in
adjacent tracks as a result of collision between them (a), and its
subsequent spontaneous rearrangement into a stable asymmetric BS (b). The
initial norm and velocities of the solitons are $N = 2\protect\pi$ and
$v=\pm 0.1$, and $\protect\varepsilon = 3$. In the 3D case, the scenario is
quite similar.}
\label{f4}
\end{figure}

\begin{figure}[htbp]
\centerline{\includegraphics[width=8.0cm,height=4.0cm,clip]{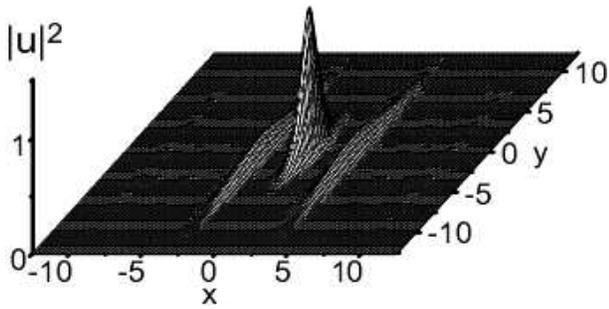}}
\vspace*{0.5cm}
\caption{Bound state of three solitons developed from the VK - unstable 2D
soliton with parameters $A=0.63, \ a = b = 0.1$, and $\varepsilon = 10$.
Half of the initial norm is lost during the formation.}
\label{f5}
\end{figure}

In the low-dimensional potential, collisions are also possible between
solitons moving in adjacent \textquotedblleft tracks\textquotedblright\
(channels). In that case, they pass each other quasi-elastically if the
collision velocity is large. If it is not too large, each soliton captures a
small \textquotedblleft satellite\textquotedblright\ in the other channel.
If the velocity falls below a critical value for
given $\varepsilon $, the outcome of the collision is altogether different:
two solitons form a quiescent symmetric bound state (BS) [Fig. \ref{f4}(a)].
Later, the symmetric BS develops an intrinsic instability, spontaneously
rearranging itself into a completely stable asymmetric BS
[Fig. \ref{f4}(b)]. This behavior resembles the
symmetry-breaking instability of optical
solitons in dual-core fibers \cite{pare} and the macroscopic
quantum-self-trapping effect in BECs confined to a double-well potential
\cite{smerzi}. In fact, this is the first example of a stable BS of two
solitons in BECs. Formation of more complex stable 2D and 3D BSs also occurs
in a different way: VK-{\em unstable} waveforms predicted by the VA undergo
violent evolution, shedding $\simeq 50\%$ of their norm and eventually
forming a BS of three solitons, with a narrow tall one in the middle, and
small-amplitude broad satellites in the adjacent channels [see Fig. \ref{f5}].

{\it Conclusion.} We have demonstrated that the periodic potential whose
dimension is smaller by $1$ than the full spatial dimension readily supports
stable single- and multi-peaked solitons in 2D and 3D self-focusing media
(although the quasi-1D potential cannot stabilize 3D solitons), which
suggest new settings for experimental search of solitons. 2D solitons exist
in a finite interval of the values of the norm, and 3D ones are stable if
the lattice strength exceeds the minimum value. In the case of
self-repulsion with a parabolic trap, stable localized states are found too.
The dependence of their structure on the lattice strength is opposite to
that in the case of the self-attraction: with the increase of the strength,
a single-peaked state is changed by a multi-peaked one.

These solitons are the first example of {\em \ mobile} multi-dimensional
pulses in BECs. Head-on collisions between the in-phase solitons may lead to
their fusion and collapse, while out-of-phase solitons collide elastically
indefinitely many times. Collision between solitons in adjacent tracks may
create a bound state (BS), which then relaxes to an asymmetric stable shape.
Three-soliton BSs are created as a result of the evolution of initially
unstable solitons.

{\it Acknowledgements.} We appreciate discussions with Y.S. Kivshar. B.B.B.
thanks the Physics Department at the University of Salerno (Italy) for a
research grant. B.A.M. appreciates hospitality of the same Department, and
partial financial support through the Excellence-Research-Center grant No.
8006/03 from the Israel Science Foundation. M.S. acknowledges partial
financial support from the MIUR, through the inter-university project
PRIN-2000, and from the European LOCNET grant HPRN-CT-1999-00163.

\end{multicols}

\end{document}